\documentclass[graybox,natbib,nosecnum]{svmult}
\bibpunct{(}{)}{;}{a}{}{,} 

\usepackage{mathptmx}       
\usepackage{helvet}         
\usepackage{courier}        
\usepackage{type1cm}        

\usepackage{makeidx}         
\usepackage{graphicx}        
\usepackage{multicol}        
\usepackage[bottom]{footmisc}
\usepackage[normalem]{ulem}	
\usepackage{hyperref}  

\usepackage{soul}   
 \hypersetup{pdftitle={Exoplanets with SOFIA},pdfauthor={D.Angerhausen}}
\usepackage{sidecap}

\newcommand{\hbindex}[1]{\hl{#1}\index{#1}}  

\makeindex             

\begin{document}

\title*{Exoplanet Research with the Stratospheric Observatory for Infrared Astronomy (SOFIA)}
\titlerunning{Exoplanets with SOFIA} 
\author{Daniel Angerhausen}
\institute{Daniel Angerhausen \at Center for Space and Habitability, University of Bern, Sidlerstrasse 5, 3012 Bern, Switzerland  \email{daniel.angerhausen@csh.unibe.ch}
}
%
%
\maketitle

\abstract{When the  Stratospheric Observatory for Infrared Astronomy (\hbindex{SOFIA}) was conceived and its first science cases defined, exoplanets had not been detected. Later studies, however, showed that \hbindex{optical} and \hbindex{near-infrared} (NIR) photometric and spectrophotometric follow-up observations during planetary transits and eclipses are feasible with SOFIA's instrumentation, in particular with the HIPO-FLITECAM and FPI+ optical and near-infrared instruments. Additionally, the airborne-based platform SOFIA has a number of unique advantages when compared to other ground- and space-based observatories in this field of research. Here we will outline these theoretical advantages, present some sample science cases and the results of two observations from SOFIA's first five observation cycles -- an observation of the Hot Jupiter HD 189733b with HIPO and an observation of the Super-Earth GJ 1214b with FLIPO and FPI+. Based on these early products available to this science case, we evaluate SOFIA's potential and future perspectives in the field of optical and infrared exoplanet \hbindex{spectrophotometry} in the \hbindex{stratosphere}.}

\section{Introduction }

\subsection{SOFIA}

\begin{figure*}
      \centering
      \includegraphics[width=0.95\textwidth]{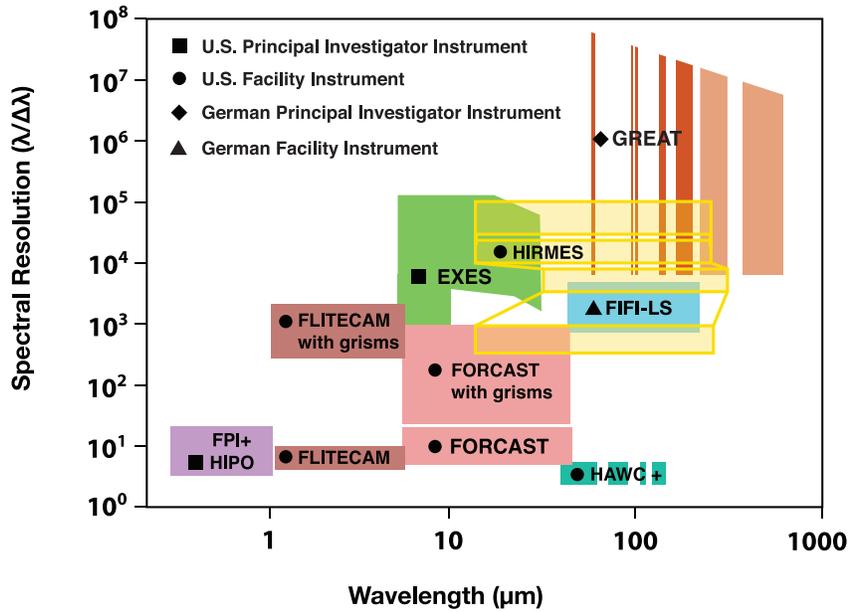}
      \caption{SOFIA's instrument suite.}
      \label{fig:inst}
    \end{figure*}

The Stratospheric Observatory for Infrared Astronomy (SOFIA), a joint US/German project of the
National Aeronautics and Space Administration (\hbindex{NASA}, 80\%) and the German Aerospace Center (\hbindex{DLR}, 20 \%), is a 2.5-m telescope on board a Boeing
747-SP aircraft. It was engineered to conduct astronomical observations in the wavelength regime between
0.3 $\mathrm{\mu m}$ and 1.6 mm (see Figure \ref{fig:inst}), while flying as high as 41,000 to 45,000 feet (12-13.7 km), above 99.8\% of the obscuring
atmospheric water vapor (Gehrz et al. 2010). The SOFIA aircraft is stationed at
NASA’s Armstrong Flight Research Center in Palmdale, CA., but also operates
from other bases worldwide (such as Christchurch, NZ during northern summer deployments) to enable observations at any declination and to facilitate timely observations of transient events, including planetary occultations and extrasolar planet transits.
The image quality currently obtained in SOFIA observations is about 2.5 arcsec FWHM, mainly due to jitter from the telescope caused by the air flow into the cavity. Shear layer seeing (air flow over the cavity) is negligible at the longer MIR and FIR wavelengths, but can be up to 5 arcsec in the optical and less in the near-infrared. Pointing of the SOFIA telescope is accurate to about 0.5 arcsec and the tracking is good, with about 0.5 arcsec over a timespan of half an hour to an hour \citep{ASNA:ASNA201311908}.
SOFIA’s first generation science instruments consist of high speed photometers, imaging
cameras, and spectrographs capable of resolving both broad features due to dust and large
molecules, and molecular and atomic gas lines at km/s resolution. SOFIA 's prime science focus is star formation and interstellar medium observations. It is a key facility for studying thermodynamics and energetics in regions of star formation (e.g.cloud collapse) as well as astro-chemical processes in the interstellar medium (ISM), including molecular rotational excitation. 

SOFIA is designed to
incorporate instrumentation upgrades in response to new technological developments much faster and easier than space-based platforms. Preliminary science observations began in December 2010, with a limited 
instrument complement and number of science flights.  Full science operations, 
when the Observatory entered Phase E, began in May 2014, with long term plans 
to sustain the Observatory for 20 years of operations (through May 2034). 
SOFIA's prime mission duration is 5 years, after which the program  will 
undergo reviews for an extended mission per NASA's Science Mission Directorate 
Senior Review process. Detailed descriptions of
the scientific and operational advantages of the SOFIA observatory, its development and
deployment schedule, and opportunities for participation by observers and instrument
developers are described for example in \cite{2009SPIE.7453E..02B}, \cite{2009AdSpR..44..413G} and \cite{2014SPIE.9145E..0QY}.

\subsection{Exoplanets with SOFIA}

Studying extrasolar planets is one of the major frontiers of present-day astronomy and astrophysics. The field has
transformed from discovery of individual extreme exoplanetary systems towards a comprehensive categorization and characterization of exoplanet populations. One important science case in this context is the characterization of exoplanetary atmospheres with observations in planetary \hbindex{transits}, eclipses and phase-curves.

\begin{figure*}
      \centering
      \includegraphics[width=0.95\textwidth]{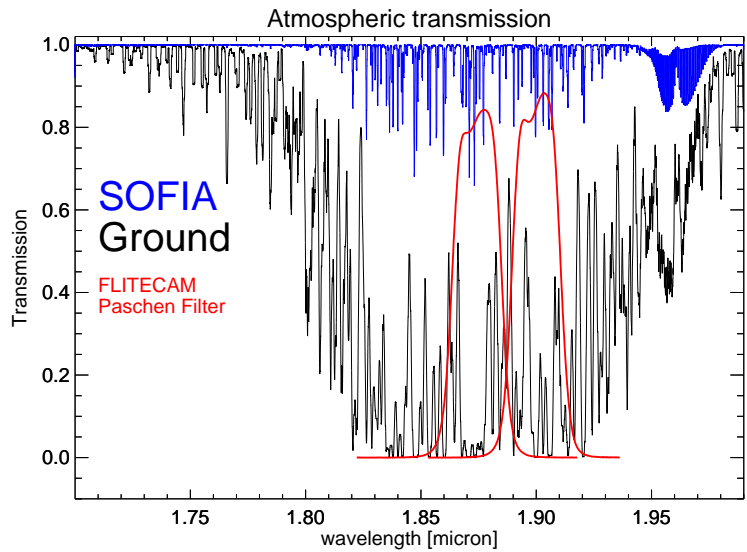}
      \caption{Transmittance of Earth's atmosphere between H and K band in the NIR, illustrating
the importance of SOFIA for this wavelength regime. The NIR water bands become almost
opaque and highly variable at even the best observing sites from the ground, but are nearly
transparent at SOFIA altitudes. The described FLITECAM observations probed this region directly in the Paschen filter.}
      \label{fig:trans_plot}
    \end{figure*}

It has been shown that in theory SOFIA has some distinct advantages for extremely precise time-domain spectrophotometric
observations at optical and  IR wavelengths  
\citep{2010PhDT.......210A,2010PASP..122.1020A,2014IAUS..293..435A,2017arXiv170807033A}.
SOFIA operates in important wavelength regions, where the planet's black-body temperature
peaks and contrast ratios between the star and planet improve. Many
important atmospheric properties, such as the chemical constituents or the temperature-pressure
profiles of planets, can be analyzed with IR spectra, observed during transits or
eclipses of the planet.
The variability of Earth’s atmospheric transmission and emission as well as the temporal
variability of its trace gases, however, is the most crucial challenge for ground-based transit
observations. This is particularly problematic when it comes to the spectroscopic analysis of molecular features
in the exoplanet’s atmosphere that are also present as telluric trace gases (Fig. \ref{fig:trans_plot}).
SOFIA flies
high enough to be independent of near surface processes affecting the $H_2O$, $CH_4$ , and
other diagnostic lines.
The SOFIA telescope operates at much lower temperatures (240 K) than ground-based
telescopes. Therefore thermal background contributions, that are the dominant noise
source for transit observations at wavelengths longer than 3 $\mathrm{\mu m}$, are significantly
reduced.
SOFIA can observe time-critical events, such as the rare transits of long-period planets,
under optimized conditions. This was demonstrated by the SOFIA team in an observation of a Pluto occultation in
June 2011 \citep{2013AJ....146...83P}. For short-period close-in transiting planets with transits
and eclipses occurring every 1-5 days, the optimal observing schedules for ground-based
transit observations are reduced to only a few nights per year for a given observing site as the event is best observed close to target culmination and local midnight. The Hubble Space Telescope (HST),
on the other hand, is able to observe transits at many more opportunities but is in
most cases limited to series of 96 minute on/off-target batches due to its low-earth orbit.
In theory the analysis of potential flight schedules shows
that the mobile platform SOFIA is able to take off close to the optimal geographic
location for each of those events. In these scenarios the \hbindex{airborne} observatory is able to observe the
complete event continuously and with a very stable setup (telescope elevation, airmass,
etc.) during observing flights of up to 10 hours duration.

In the next subsection we discuss the available \hbindex{instrumentation} for spectrophotometry on SOFIA, in the section after that we introduce and describe accepted proposals and conducted observations with SOFIA and in the final section  we summarize our findings, present constraints of SOFIA and give an outlook on SOFIA's future in this field.

\subsection{SOFIA's Instruments for optical and NIR spectrophotometry}

\textbf{\hbindex{HIPO.}} The \textit{High Speed Imaging Photometer for Occultations} (HIPO) is a Special Purpose Principal Investigator class Science Instrument (SSI, \cite{dunham04, 2014SPIE.9147E..0HD}).  HIPO is designed to provide simultaneous high-speed time resolved imaging photometry at two optical wavelengths. The primary HIPO detectors are e2v CCD47-20 $1024\times 1024$ pixel frame transfer CCDs with plate scales of $0.33''\times 0.33''$ pixels at low resolution and $0.05''\times 0.05''$ pixels at high resolution. The HIPO field of view (FoV) is a 5.6' square, the 8' diagonal of which corresponds to the 8' diameter SOFIA field of view. The filter set includes the Johnson (UBVRI) and Sloan (u'g'r'i'z') filters as well as a filter for methane at 890 nm. A number of readout modes are available allowing the observer to optimize the subframe size, speed, noise, full well capacity and linearity trade off for any particular event. The duty cycle for both HIPO channels is essentially 100 percent with a minimal dead time due to the CCD frame transfer of a few milliseconds, because in the frame transfer CCD the readout is overlapped with the next integration. HIPO is no longer offered through the normal calls for proposal, either stand-alone or in combination with FLITECAM. It is, however, still available through use of the remaining Guaranteed Time Observations assignment of the PI.

\textbf{\hbindex{FLITECAM.}}
 The  \textit{First Light Infrared TEst CAMera} (FLITECAM) is a near-infrared imager and grism spectrograph covering the $1 - 5 \mu$m range (\cite{mclean06, smith08, logsdon14}).  FLITECAM's $1024\times 1024$ InSb ALADDIN III array covers an $\sim 8$ arcmin diameter field of view with a plate scale of 0.475 arcsec  pixel$^{-1}$. The full set of available FLITECAM filter passbands are listed on-line in the FLITECAM Observers Handbook. FLITECAM can be co-mounted with the HIPO instrument during transit observations, a configuration that precludes observations at wavelengths longer than $\sim 4~\mathrm{\mu m}$, though, and that reduces the sensitivity at wavelengths longer than $\sim 2~\mathrm{\mu m}$, due to high background levels resulting from the warm dichroic and transfer optics. FLITECAM, either stand-alone or co-mounted with HIPO, is not being offered for observations in the current cycle 6 of SOFIA.  Limited new observations may be requested through the Director's Discretionary Time process. 

\textbf{\hbindex{FPI+.}} On SOFIA the light is passed through the telescope dichroic tertiary mirror (25\% and 45\% reflectivity for the B and z' bandpasses) to the \textit{Focal Plane Imager }(FPI+, \cite{2016SPIE.9908E..2WP}). The FPI+ contains a highly sensitive and fast EM-CCD camera. Its images are primarily used for tracking but can also be stored without disrupting the tracking process and in parallel with measurements of the instruments mounted to the telescope. With the released call for proposals for the SOFIA observing cycle 4 (2015), the FPI+ was made available for proposals as a facility science instrument for observations in 2016 and after.

\section{Spectrophotometric Exoplanet observations with SOFIA}

\subsection{Accepted proposals in cycles 1-4}

In Table \ref{tab:prop} we give an overview of the accepted proposals in the area of exoplanet spectrophotometry. In this section we briefly summarize these proposal and then present the observations and results that have been published to date. 

\begin{table}[ht!]
\centering
     \caption[]{Overview: accepted exoplanet proposals for SOFIA}
    \label{tab:prop}
      \begin{tabular}{l |l| l| l}
      \hline\hline
      \noalign{\smallskip}
Prop. ID  & Title & PI &  Results\\
    \hline\hline
		\parbox[t]{1cm}{01 0099} & \parbox[t]{4cm}{Characterizing Transiting Exoplanets Using FLITECAM: An Exploratory Program} & \parbox[t]{2cm}{ \cite{2012sofi.prop...99M} }&   \parbox[t]{4cm}{Failure of FPI and FLITECAM, two HIPO channels published in   \cite{2015JATIS...1c4002A} } \\
        \hline

		\parbox[t]{1cm}{02 0053 02 0084} &  \parbox[t]{4cm}{Exoplanet transits with FLIPO: Is GJ 1214b a water-world Super Earth or a cloudy Mini-Neptune?} &  \parbox[t]{2.5cm}{\cite{2013sofi.prop...53A} \\ \cite{2013sofi.prop...84D} }&   \parbox[t]{4cm}{Observation successful, published in \cite{2017arXiv170807033A} } \\
        \hline

		\parbox[t]{1cm}{01 0155  02 0046 03 0042} &  \parbox[t]{4cm}{Do starspots inflate the exoplanet CoRoT-2b?} &  \parbox[t]{2.5cm} {\cite{2012sofi.prop..155H} \\ \cite{2013sofi.prop...46H} \\ \cite{2014sofi.prop...42H}}&   \parbox[t]{4cm}{Observed in October 2016, observation are strongly affected by engine glint problem } \\
        \hline

        	\parbox[t]{1cm}{03 0037} &  \parbox[t]{4cm}{The Origin of non-LTE Emission on Dayside of a hot-Jupiter Exoplanet} &  \parbox[t]{2.5cm} {  \cite{2014sofi.prop...37S}}&   \parbox[t]{4cm}{Succesfully observed in October 2015, results in preparation (Swain, priv. comm) } \\
        \hline    
        
\parbox[t]{1cm}{03 0052} &  \parbox[t]{4cm}{Observation of the primary transit of GJ 3470b: Warm Uranus transmission spectrophotometry with FLIPO} &  \parbox[t]{2.5cm} {  \cite{2014sofi.prop...52A}}&   \parbox[t]{4cm}{Observed in September 2015 without HIPO (only FPI and FLITECAM), data in preparation, some technical issues } \\
        \hline    
	    \noalign{\smallskip}
        
        \parbox[t]{1cm}{04 0024} &  \parbox[t]{4cm}{Seeing SPOTS with SOFIA: Starspot Photometric Observations of Transiting Systems} &  \parbox[t]{2.5cm} {  \cite{2015sofi.prop...24G}}&   \parbox[t]{4cm}{Not observed } \\
        \hline    
	    \noalign{\smallskip}
      
      \end{tabular}
\end{table}

In their proposal \textit{Characterizing Transiting Exoplanets Using FLITECAM: An Exploratory Program} \cite{2012sofi.prop...99M} and collaborators proposed to explore SOFIA's exoplanet characterization capability by conducting observations of two transiting exoplanets, \hbindex{HD 189733 b} and WASP-12 b, using both the photometry and spectroscopy modes of FLITECAM to measure water absorption at 1.85 $\mathrm{\mu m}$. These two exoplanets presented scientifically compelling cases for preliminary observations, while also providing the opportunity for obtaining high-precision data required to pave the way for future observations of a wide range of exoplanet atmospheres with SOFIA/FLITECAM. Some parts of this program were observed and the results of this observation are published in \cite{2015JATIS...1c4002A} and are summarized below.

In their joint US-German proposal\textit{ Exoplanet transits with FLIPO: Is \hbindex{GJ 1214b} a water-world Super Earth or a cloudy Mini-Neptune?} PIs \cite{2013sofi.prop...53A} and  \cite{2013sofi.prop...84D} proposed to use FLIPO on SOFIA to comprehensively analyze the atmospheric composition and temperature structure of the  super-Earth planet and possible water-world GJ 1214b.  Constraints on the planetary atmosphere would be achieved through spectrophotometric transmission observations during transit in 2 optical and 2 infrared channels - the 1.90 $\mathrm{\mu m}$ `Paschen alpha cont.' and 3.05 $\mathrm{\mu m}$ `water ice' bands - that are not observable from the ground. Again only parts of the full program were observed and are published in \cite{2017arXiv170807033A}  and summarized below.

In their cycle 3 proposal 
\textit{Observation of the primary transit of GJ 3470b: Warm Uranus transmission spectrophotometry with FLIPO} PI \cite{2014sofi.prop...52A} and collaborators proposed to probe the atmosphere of \hbindex{GJ 3470b} measuring the planet's radii during one transit, in three optical and in one infrared channels with HIPO and FLITECAM (in the FLIPO configuration with one additional channel from the FPI+) on SOFIA. These observations were conducted, but only with FPI+ and FLITECAM and suffered from some technical issues. The data is currently analyzed in order to reveal if these issues can be corrected for in the reduction process.

For the other cycle 3 proposal \textit{The Origin of non-LTE Emission on Dayside of a hot-Jupiter Exoplanet
} PI \cite{2014sofi.prop...37S} and collaborators
proposed to use SOFIA/FLITECAM in spectroscopic mode to observe a secondary eclipse of the Hot Jupiter HD 189733b in order to decisively confirm and probe the physical origin of a previously reported strong emission around 3.2 $\mathrm{\mu m}$ \citep{2010Natur.463..637S}. This non-LTE (Local Thermodynamic Equilibrium) emission is attributed to methane and may originate from high J number hot combination bands, but improved measurements are needed to confirm this. This observation was conducted and the publication of the results currently is in preparation \citep{2016AAS...22732106Z}.

The proposal \textit{Do starspots inflate the exoplanet CoRoT-2b?} by PI \citep{2013sofi.prop...46H} and collaborators has been selected on the German TAC in cycle 1, 2 and 3. They proposed to use SOFIA’s combination of HIPO and FLITECAM to obtain simultaneous optical and infrared transit photometry of the CoRoT-2 system. This system consists of the highly active planet host-star CoRoT-2A and the unusually inflated hot Jupiter \hbindex{CoRoT-2b}. CoRoT-2A’s surface is densely covered with \hbindex{starspots}, which influence the transit lightcurves and, thus, complicate the determination of accurate planetary parameters. In particular, this can lead to an overestimation of the planetary radius, which could, at least partially, account for the unusually large radius inferred for CoRoT-2b. Using SOFIA, the weaker starspot contrast at longer wavelengths would allow  to derive an accurate planetary radius from infrared photometry and, thereby, pin down CoRoT-2b’s radius anomaly. Because of SOFIA’s unique capability to simultaneously obtain optical photometry, the data would also allow to determine starspot temperatures and estimate the total spot-coverage of CoRoT-2A’s surface by comparing the deformation and depth of the infrared and visual-band transit lightcurves. Only SOFIA’s combination of HIPO and FLITECAM provides the quasi space-based high-precision multi-band photometry required to address these open questions on the young planetary system CoRoT-2. This observation was finally conducted in September 2015 but suffered from an engine glint problem (see below).

 In a very similar proposal \textit{Seeing SPOTS with SOFIA: Starspot Photometric Observations of Transiting Systems}
PI \citep{2015sofi.prop...24G} and collaborators 
proposed to utilize the unique capabilities provided by HIPO/FLITECAM (FLIPO) in combination with FPI to obtain simultaneous, time-resolved multi-color photometry extending from the visible to the near infrared of magnetically active dwarf stars that are also the hosts of transiting exoplanets. In this way, they aim to measure the fundamental properties of starspots that are eclipsed by the transiting planet and thereby provide much improved constraints for starspot models than has been possible with the single-band data from Kepler and CoRoT. Since starspots are the strongest concentrations of magnetic flux on the Sun and stars, an understanding of their properties can yield critical constraints for stellar dynamo models and influence our views of the role of stellar magnetic activity in star-planet interactions.

\subsection{Observational parameters in the airborne environment, noise models and light-curve fitting} \label{sec:obsparam}
Photometric observations from an airborne platform like SOFIA differ from ground-based observations. While ground-based photometry suffers from systematic errors induced by e.g. air mass or local weather changes, photometric observations with SOFIA also correlate with changes in flight parameters such as Mach-number or air density. 

\begin{figure*}
      \centering
      \includegraphics[width=0.95\textwidth]{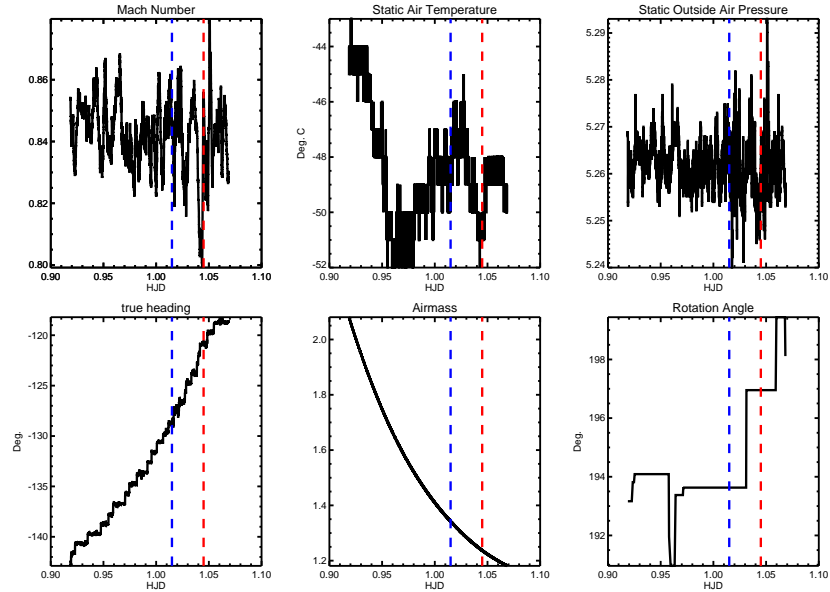}
      \caption{Sample time series of some observational parameters fo the GJ 1214b observation to illustrate the mobile airborne environment. The beginning of ingress and end of egress are marked in blue and red vertical lines. Adapted from \cite{2017arXiv170807033A} }
      \label{fig:hk_plot}
    \end{figure*}

For both published SOFIA observations presented later in this section the authors used the housekeeping data taken during the observations to parameterize the time dependence of the observational environment. Figure \ref{fig:hk_plot} shows the time series of selected observational parameters, some of them unique to the airborne environment. Many of these parameters are mutually correlated. In order to overcome these degeneracies they performed a principal component analysis on all available parameters to produce a set of linearly independent time series to de-correlate the raw light curves.

For these observations the noise was modeled with the first 16 principal components $p_i(t)$ in order to de-correlate with the observational parameters, sampled at the same time as their exposures, as a linear combination $R_{model}(t)=\sum c_i \times p_i(t)$.

In \cite{2015JATIS...1c4002A}, the resulting light-curves were then fitted using the Transit Analysis Package (TAP) \citep{Gazak12} which is  built on EXOFAST \citep{eastman13}. In \cite{2017arXiv170807033A}  they used
the `Transit Light Curve Modeller (TCLM)' code \citep{csizmadia11, csizmadia15} and the 
Bayesian model selection, and parameter estimation code EXONEST \cite{placek+14, placekknuth15,placek+15}.
Their final results are summarized in the following section.

\subsection{Published Results}\label{sec:pubres}

 \subsubsection{HD 189733b with HIPO}  
 
 \cite{2015JATIS...1c4002A} observed HD 189733 b during a transit on SOFIA's flight  number 134 on UT Oct 1 2013 as part of the Cycle 1 GO program (PI: Mandell, Proposal ID: 01-0099, see Table \ref{tab:prop}).  These observations were conducted in the FLIPO configuration (FLITECAM and HIPO operating simultaneously) with the goal to observe in three optical and infrared bands at the same time: B and $z^\prime$ with HIPO and a narrow-band filter covering the Paschen $\alpha$ spectral feature at 1.88 $\mathrm{\mu m}$ with  FLITECAM. The HIPO filters were  selected to avoid spectral regions with potentially high ozone variability, while the FLITECAM filter was chosen due to its wavelength coverage of a prominent $H_2O$ spectral feature that cannot be sampled from ground-based observatories. An additional optical channel (for general calibration or tracing of a specific telluric absorption band) was planned to be obtained from the Focal Plane Imager (FPI+). However, due to an instrument malfunction they were unable to acquire FPI+ and FLITECAM data for this flight.

Due to constraints on the flight plan imposed by requirements on the direction and timing of SOFIA flights (see constraints section), the observing period only allowed for a very short baseline before ingress and almost no baseline after egress. 
 
\begin{figure*}
      \centering
      \includegraphics[width=0.95\textwidth]{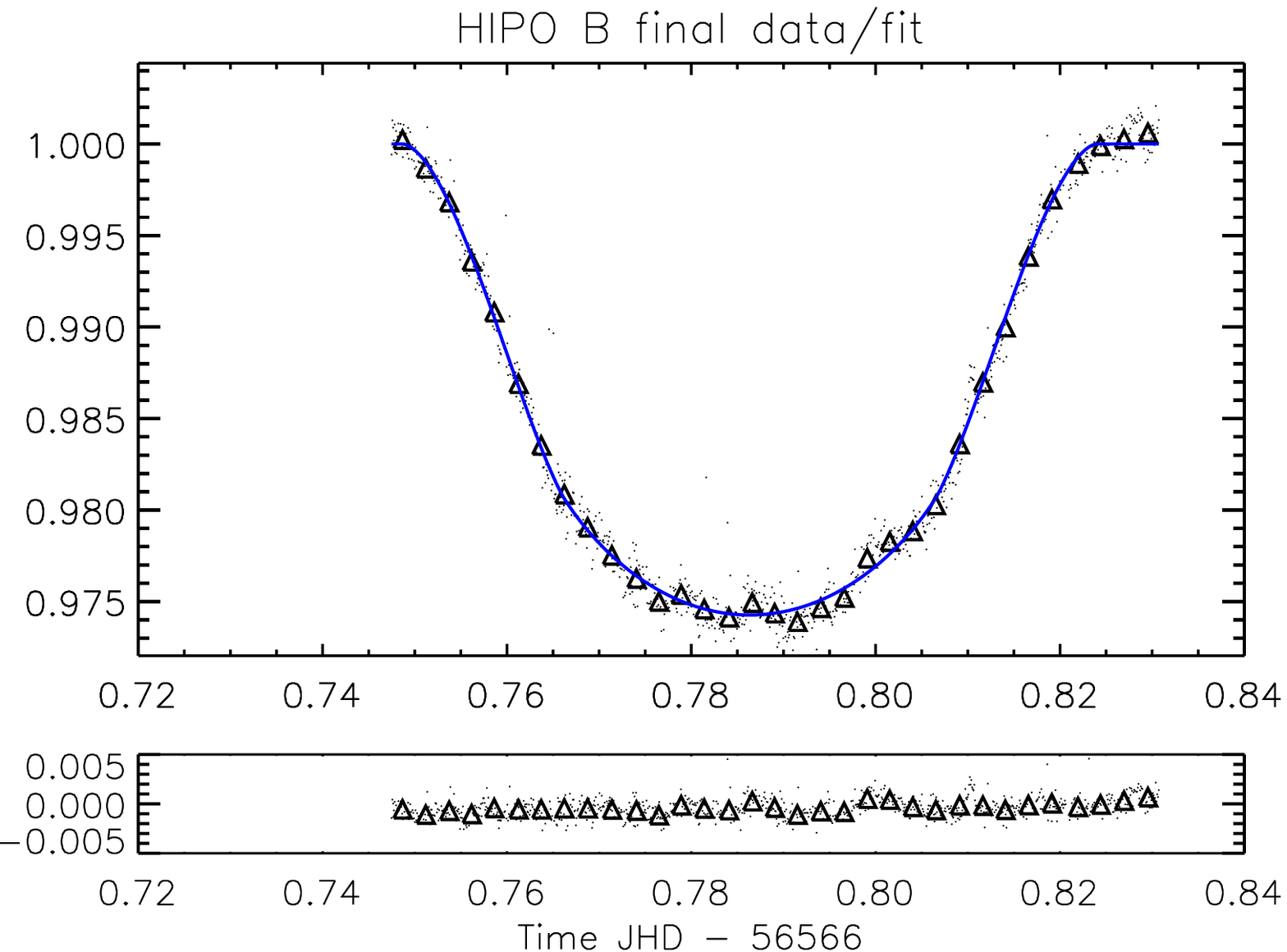}
      \includegraphics[width=0.95\textwidth]{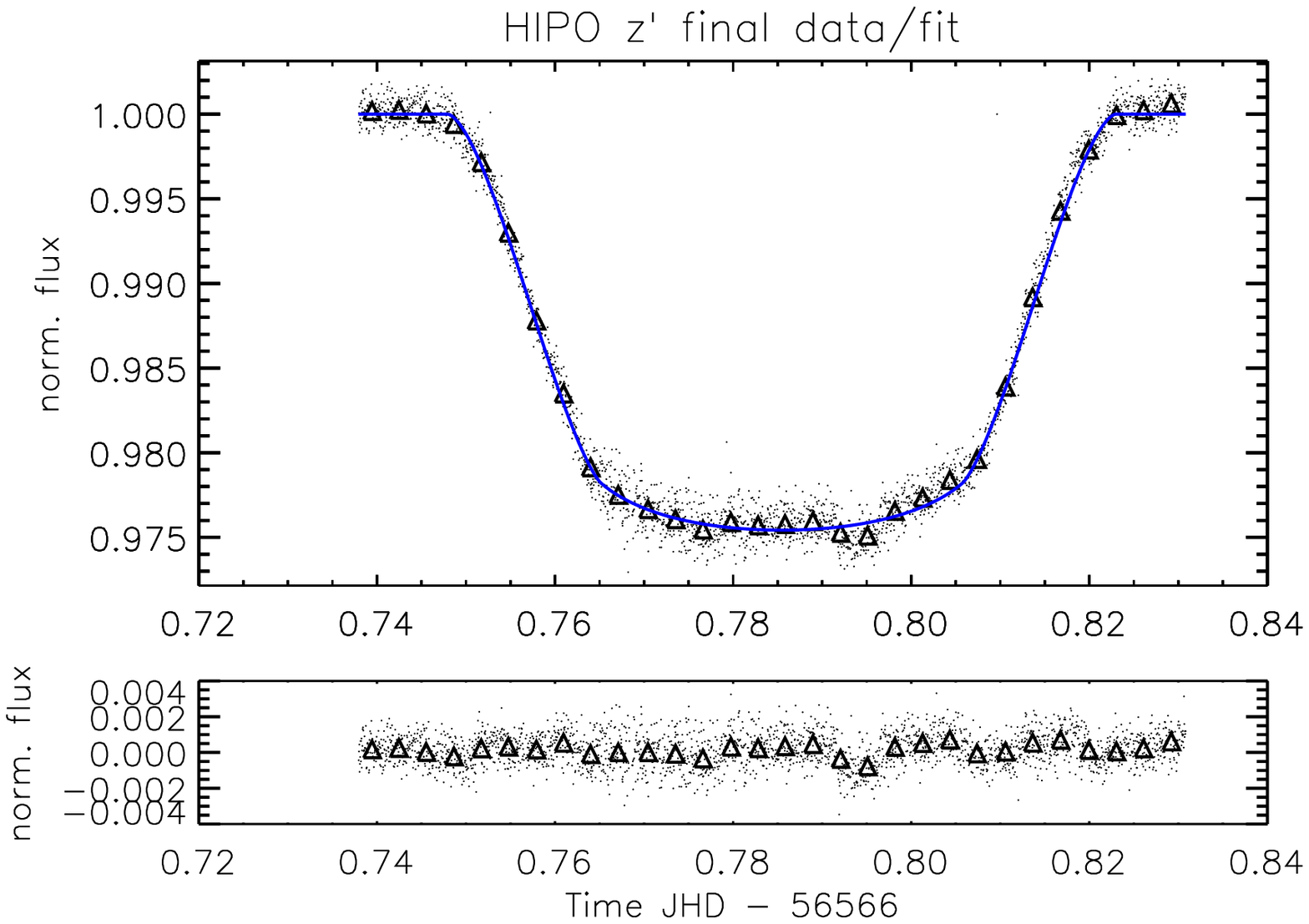}
      \caption{Final light-curves and fits for the very first exoplanet transit observation with SOFIA on October 1 2013. Plots show the HIPO blue (top) and z' band (bottom) transits of HD~18933b. From \cite{2015JATIS...1c4002A}.}
      \label{fig:189_lc}
    \end{figure*}

In  \cite{2015JATIS...1c4002A} they summarize this first successful exoplanet transit observation SOFIA as a demonstration of its capability to perform absolute transit photometry, i.e. observing transit light-curves without the need of a comparison star (see Figure \ref{fig:189_lc}). They present a detailed description of our data reduction, in particular, the correlation of \hbindex{photometric systematics} with various in-flight parameters unique to the airborne observing environment, which we summarized in the previous section. The derived transit depths from this observation at B and z' wavelengths confirm a previously reported slope in the optical transmission spectrum of HD 189733 b. Their results gave new insights to the discussion about the source of this \hbindex{Rayleigh scattering} in the upper atmosphere and the question of fixed limb darkening coefficients in fitting routines.
  
  They argue that accounting for the fact that the observation (1) had very little out of transit baseline, (2) was taken with an only $30\%$ reflective tertiary and (3) assuming there is still room for improvement for the correction of systematics once they are better understood with further observations, they could report that SOFIA was able to deliver space based data quality in an airborne environment at least in the optical with HIPO. 

          \begin{figure*}
      \centering
      \includegraphics[width=0.95\textwidth]{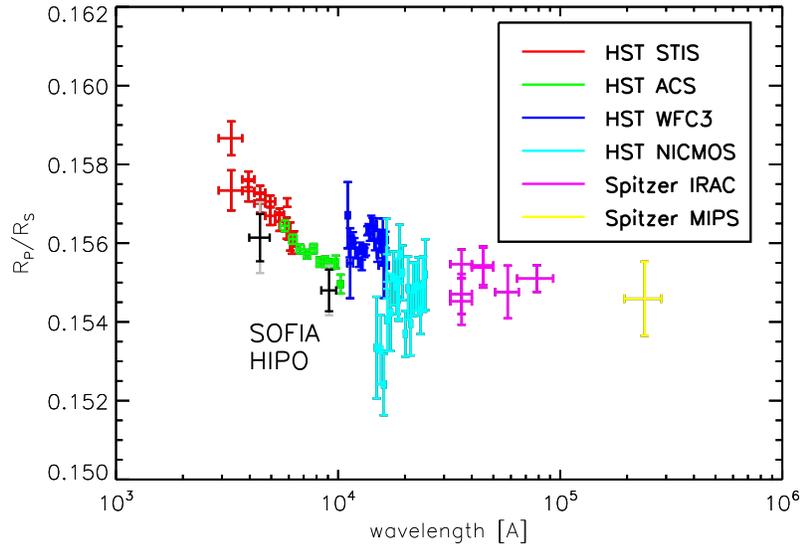}
      \caption{Final results of the HIPO-SOFIA HD 189733b observation (black: free limb darkening, grey: fixed limb darkening) in comparison to previous observations from other space-based platforms such as \textit{HST} and \textit{Spitzer} \citep[see legend, from ][]{2013MNRAS.432.2917P}.  Our B and $z^\prime$ data points reproduce the slope in the optical, most likely caused by Rayleigh scattering, though with an overall smaller transit depth. From \cite{2015JATIS...1c4002A}. }
      \label{fig:189_sp}
    \end{figure*}

  \subsubsection{GJ 1214b with FLIPO and FPI+}   
  
  The observation proposed in the joint US-German Cycle 2 GI program (US-proposal: \cite{2013sofi.prop...53A} ; German-proposal:  \cite{2013sofi.prop...84D}, see Table \ref{tab:prop}) was performed on SOFIA’s flight number 149 on UT February 27, 2014.

\cite{2017arXiv170807033A} observed a transit of  GJ 1214b using the photometry mode of FLITECAM and HIPO in the `FLIPO' configuration. GJ 1214b was monitored during one 52 min transit plus ca. 70 min before and 10 min after transit for a total of 150 min (including some additional time for setups and calibrations). Observations were simultaneously conducted in two optical HIPO channels: open blue at $0.3 - 0.6\,\mathrm{\mu m}$ and Sloan z' at $0.9\,\mathrm{\mu m}$ and one infrared FLITECAM filter: Paschen-$\alpha$ cont. at $1.9\,\mathrm{\mu m}$. Complementary data were also obtained with the optical focal plane guiding camera FPI+ in the Sloan i' band ($0,8\,\mathrm{\mu m}$), as it was used for both tracking and data acquisition purposes. Due to flight planning constraints, the end of the transit occurred in morning twilight and in the last $\sim$15 minutes the sky brightness gradually increased to about 3.5 times its night-time values. Furthermore their measurements suffered from insufficient calibration files for the FLITECAM channel (see below).

\begin{SCfigure*}
      \centering
      \includegraphics[width=0.65\textwidth]{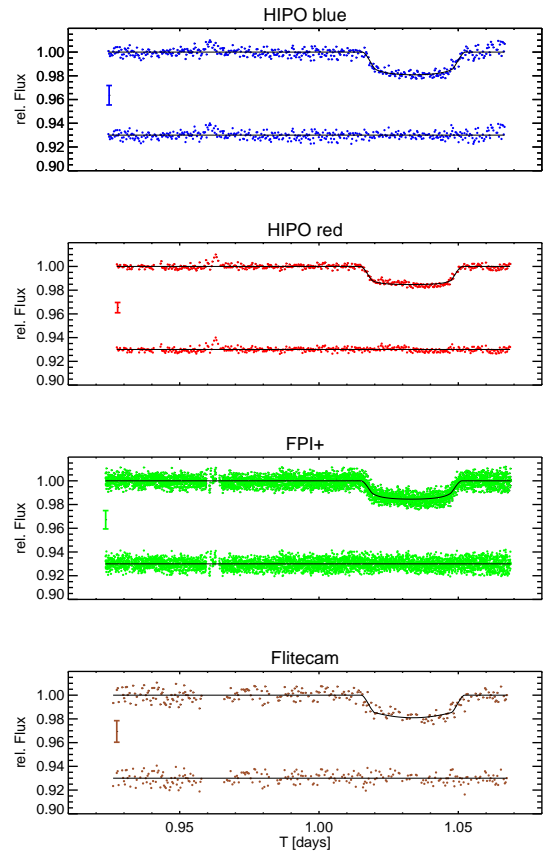}
      \caption{Final light-curves and fits for the GJ 1214b transit observation with SOFIA. Plots show the HIPO open blue (top) and z' band, FPI+ I-band and FLITECAM Paschen alpha (bottom) transits of GJ 1214b. From \cite{2017arXiv170807033A}.}
      \label{fig:1214_lc}
    \end{SCfigure*}

\begin{figure*}
      \centering
      \includegraphics[width=0.95\textwidth]{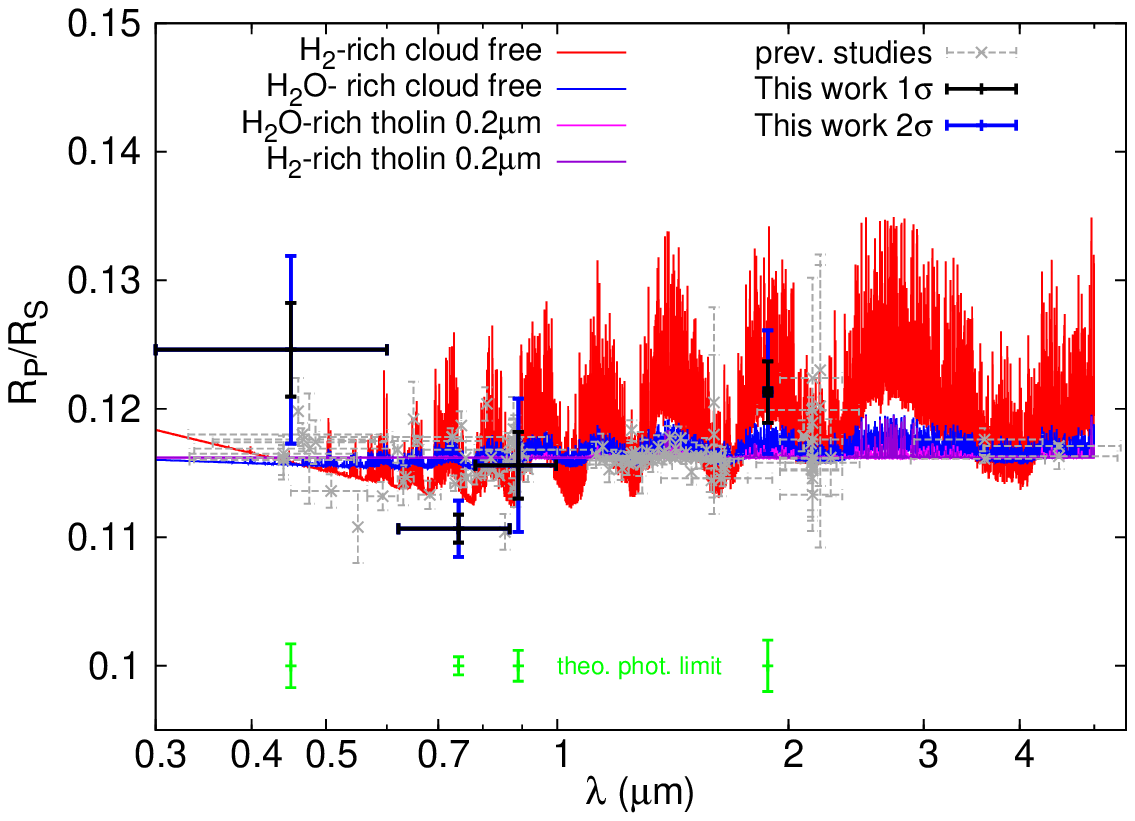}
      \caption{Final spectrum of GJ 1214b. From \cite{2017arXiv170807033A}.} 
      \label{fig:1214_sp}
    \end{figure*}

However, they were able to derive four simultaneous light curves with sensitivities between 1.5 and 2 times the photon noise limit and corresponding transit depths in three optical and one infrared channel, which they compare to previous observations and state-of-the-art synthetic atmospheric models of GJ 1214b (see Figures \ref{fig:1214_lc} and \ref{fig:1214_sp}).
This was the first (and so far only) exoplanet observation that used SOFIA's full set of instruments
available to exoplanet spectrophotometry.  Therefore they were able to evaluate SOFIA’s potential in this
field and suggest future improvements.   We summarize these finding below.

\section{Identified Problems and Constraints}\label{sec:const}

\textbf{Flat Fields.} For FLITECAM it is complicated and time consuming to obtain a reliable flat field on such narrow band filters as the 1.9 $\mathrm{\mu m}$ Paschen-$\alpha$ (continuum) filter. Building up enough sensitivity on the flat fields for ppm photometry requires very long exposures that are difficult to accommodate during a SOFIA campaign. Therefore one has to use for example K-band flat fields preferably taken on the same flight or campaign. \\

\noindent\textbf{Line-of-Sight (LOS) rewinds.} In the FLIPO setup SOFIA does not provide an image rotator to compensate field rotation during long integrations. This introduces a rotation of the images over time. Due to SOFIA's unique setup, the telescope must periodically undergo so-called `Line-of-Sight (LOS) rewinds'. The required frequency of LOS rewinds depends on rate of field rotation experienced by the target, which is a complex function of the position of the target in the sky relative to that of the aircraft heading. These need to be carefully timed with regard to the transit observation, to not interfere with e.g. ingress or egress. While its possible to keep the target star mostly in the boresight, the rest of the field moves over the CCD due to this field rotation. This is one of the main factors introducing systematic noise and limiting the photometric precision of the instrument and is another reason why reliable flat fields are crucial for this kind of time series observation. \\

\noindent\textbf{Background} With an operating temperature of about $-50^{\circ}$C SOFIA has significantly less \hbindex{thermal background} contributions than ground-based facilities. This gives SOFIA an advantage in the 3-5 $\mathrm{\mu m}$ wavelength range. Beyond 5 $\mathrm{\mu m}$ SOFIA's capabilities in exoplanet spectrophotometry are still very limited due to background contributions for all but the very brightest targets.\\

\noindent\textbf{Engine glint.} Another problem for FLITECAM is  a glint from the engine cone seen at 1.6 to 3 $\mathrm{\mu m}$.   It has been discovered to be elevation dependent and is seen below 45 deg. How important it is, will depend on the elevation of the observations.  This glint is added radiation, and can not be flat fielded out.  Based primarily on the data from the Pluto occultation flight, it was found that these features seen in FLITECAM images at low telescope elevations resulted from light from the engine cone and plume, directed into the beam in three ways: (a) shining 
directly onto the primary (scattered by dust on the mirror), (b) scattered from the edge of the secondary, and (c) scattered from the spider vanes. 
The dominant component during the Pluto occultation and the first set of exoplanet observations was clearly the light scattered from the spider 
vanes, as verified by pupil images. That component was greatly reduced (and is now almost negligible) by the installation of baffles on the spider 
vanes.  Before the CoRoT-2b observation in September 2016 the secondary mirror was re-aligned. This caused 
the scattered light component from the edge of the secondary resulting in substantially increased features in the focal plane at low telescope elevations. To fix this problem the telescope engineers have designed a baffle for the edge of the secondary that will be installed in fall 2017. This should 
eliminate this particular scattered light component (Becklin, Vacca, priv. comm.).  \\

\begin{SCfigure*}
      \centering
      \includegraphics[width=0.65\textwidth]{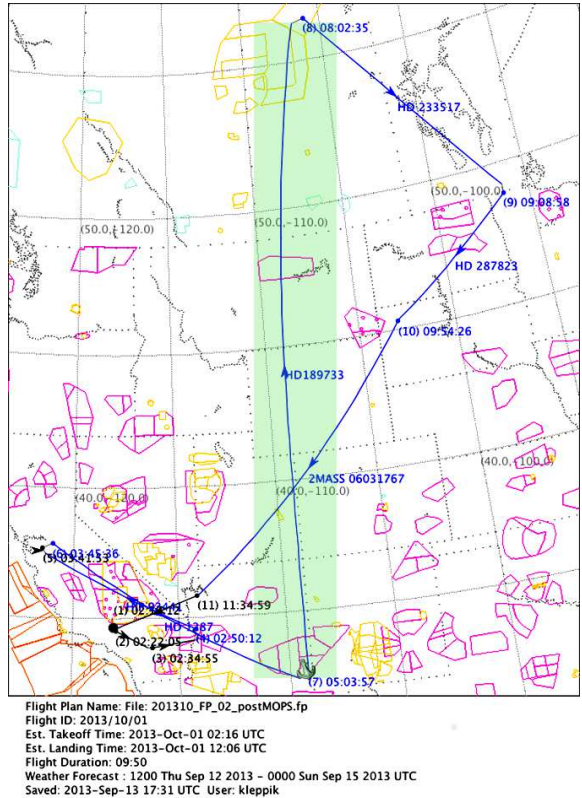}
      \caption{Flight plan for SOFIA flight 134, Oct 1 2013.  The exoplanet leg is the north-south leg in the center.}
      \label{fig:fp_plot}
    \end{SCfigure*}

\noindent\textbf{Flight planning} In theory SOFIA, as a mobile platform, is able to fly to the optimal geographic position to observe rare transient events such as exoplanet transits. In practice, however, various constraints to the flight planning make transit observations with SOFIA very complicated. The telescope has an unvignetted elevation range of 20 to 60 degrees. Since the cross-elevation range is only a few degrees, most of the azimuthal telescope movement required during tracking must be provided by changing the airplane’s heading. This requirement also dictates that the flight plan is determined by the list of targets to be observed. For flights that take off and land from the same field, some fraction of the targets must be located to the north since the telescope can only view from the portside of the aircraft. A single target can in principle be viewed for an entire flight by flying one-way flights between widely separated airfields \citep{ASNA:ASNA201311908}. However, SOFIA is usually limited to start and land in Palmdale, CA. Furthermore it has to land before sunrise to avoid accidental open door landings during daylight. In practice this limits SOFIA's ability to observe continuous legs on one single object to about 4-5 hours. As shown in Figure \ref{fig:fp_plot} SOFIA's flight plans have to avoid certain restricted flight zones and need to be optimized to accommodate all planned observation legs.

\section{Conclusions}

\subsection{Potential instrument updates for SOFIA}
As mentioned above it is possible to update SOFIA's instrumentation with a modernized VIS/NIR precision photometer similar to the previously proposed NIMBUS concept \citep{2012SPIE.8446E..7BM}. This accompanied by a reliable and robust water vapor monitoring system, can make important SOFIA-unique contributions to exoplanet science. 

 NIMBUS \citep{2012SPIE.8446E..7BM}
is an innovative near-infrared multi-band ultraprecise spectroimager (\hbindex{NIMBUS}) for SOFIA. It was designed to characterize a substantial population of extrasolar planet atmospheres by measuring atmospheric composition and dynamics during primary transit and occultation. This wide-field spectroimager could also be used for other science cases such as Trans-Neptunian Objects (TNO), Solar System occultations, brown dwarf atmospheres, carbon chemistry in globular clusters, chemical gradients in nearby galaxies, and galaxy photometric redshifts.
NIMBUS' optical design splits the light from the telescope into eight separate spectral channels, centered around key molecular bands from 1 to 4 $\mathrm{\mu m}$, that were mostly chosen in windows not observable from the ground. Each spectral channel has a wide field of view for simultaneous observations of a reference star that can decorrelate time-variable atmospheric and optical assembly effects, allowing the instrument to achieve ultraprecise calibration for imaging and photometry for a wide variety of astrophysical sources. NIMBUS produces the same data products as a low-resolution integral field spectrograph over a large spectral bandpass, but this design obviates many of the problems that preclude high-precision measurements with traditional slit and integral field spectrographs. This instrument concept is currently not funded for development. 

\subsection{Future Perspectives for Exoplanet Photometry with SOFIA}

In this review we presented the first spectrophotometric exoplanet observations with SOFIA that used all available instruments for simultaneous exoplanet transit and eclipse spectrophotometry. 

In the current configuration and with the challenges described above there are certain niches that we are able to identify with these SOFIA exoplanet observations. 
In summary this phase space is:
\begin{itemize}
    \item  bright host stars (like HD 189733b) - for which \cite{2015JATIS...1c4002A} demonstrated the ability to perform absolute optical photometry 
 	\item  short transit durations in systems like Trappist-1 
 	\item  science cases that leverage SOFIA's unique  capability to observe IR and optical light curves  simultaneously (such as starspot characterization)
     	\item science cases that complement James Webb Space Telescope (JWST) coverage and/or can be used for JWST target selection  and support
 	\item transits that are rare and/or time-critical and require a dedicated deployment 
\end{itemize}
With the upcoming TESS (Transiting Exoplanet Survey Satellite) \citep{2014SPIE.9143E..20R,sullivan15} and PLATO (PLAnetary Transits and Oscillations of stars) \citep{rauer14, 2015ApJ...810...29H} missions we will see a lot more transiting exoplanets that fall into these categories.

\begin{acknowledgement}
D.A. acknowledges the support of the Center for Space and Habitability of the University of Bern. This work has been carried out within the frame of the National Centre for Competence in Research PlanetS supported by the Swiss National Science Foundation.
\end{acknowledgement}

\bibliographystyle{spbasicHBexo}  
\bibliography{HBexoTemplateBib} 

\end{document}